# ARCHITECTING END-TO-END INFORMATION SERVICES FOR CONTINUOUS STUDENT BEHAVIOUR MANAGEMENT

## Wayne Hellmuth
School of Science and Engineering
Queensland University of Technology
QLD, Australia
Email: wayne.hellmuth@hotmail.com

## Glenn Stewart
School of Science and Engineering
Queensland University of Technology
QLD, Australia
Email: glenn.stewart@qut.edu.au

## Abstract

Data capture and use is vital for the continuous improvement of both student learning and behavior management. Previous studies on data use in the education sector have highlighted a number of problems associated with data quality and its subsequent use. These include the accuracy, consistency, completeness, and timeliness of data. Engagement issues with data have centered on the interpretation and application of the knowledge that data can provide. No study to date has investigated the link between IS design and the production of quality data that captures student progression and outcomes in either the learning or behavior management environments. This study reports on the design, development, implementation and evaluation of a novel artefact facilitating quality data for one classroom based education service: behaviour management. This study, using Design Science Research methods, shows that information systems design is a major barrier to teacher adoption and use of classroom based Information Systems.

**Keywords:** Design Science, Enterprise Architecture, Data Quality, Continuous Improvement, Education.

## 1 Introduction

The aim of this study was to further knowledge of information systems design within the education context. Previous research on data use has tended to focus on: 1. interventions relevant to data use 2; the relationship between data and aggregate outcomes, and; 3. the technical quality of the outcome measures. These studies have been inadequate in providing any advancement towards the goals of improving the practice and effects of data use. These studies do not provide insight into the (complex) mechanisms through which education initiatives influence outcomes (Coburn and Turner, 2012; Colyvas 2012; Honig and Venkateswaran 2012; Little 2012). As Coburn and Turner state, "understanding outcomes without understanding the mechanisms that produced them means that we have little insight into how to redesign data use interventions so as to increase their impact in practice" (Coburn and Turner 2012, p. 101).

This research paper uses a Design Science research (DSR) methodology to investigate the problems associated with producing and using quality data in the education sector. Typically teachers consider the classroom to be an environment where teaching and learning is sacred. There is little tolerance for "administrative tasks" that impede the primary function of teaching and learning (Earl and Katz 2002). There is, however, increasing calls, from external sources to the school, such as parents and administrators, for improvements to the quality of data that describes student progression (Bernhardt, 1998; Dembosky, et al. 2005). Although the need for improved technology is apparent, technology designs thus far have been unable to produce IT systems that facilitate the collection and use of data without distracting away from this primary classroom task of teaching and learning. As a result, the teaching profession has not been able to collect and use data in ways that further refines teaching practices for the facilitation of continual improvements of each individual student.

Design Science Research (DSR) is a problem solving paradigm with its origins in the Engineering and Science of the artificial (Simon 1996). DSR is described as a research paradigm where knowledge and understanding of design problems is gained through the building and application of IT artefacts (Hevner and Chatterjee 2010; March and Smith 1995). Through applying a DSR methodology to the problem of data quality in education this study consisted of three main areas of development including: 1. identification of why data collected in schools is of poor quality and, therefore, teacher perception that school based IS' lack utility; 2. The design of a novel mobile based application that facilitates data collection and its use in the classroom, and; 3. As a result of instantiating the artefact, develop theory about information systems design and development for future IS iterations.

The results from this study showed that the specific design of the newly instantiated artefact improved data quality and its subsequent use, thus, facilitating and enabling continuous improvement cycles to the teaching and learning process. The results, however, reveal that exogenous factors to the artefact, categorised as socio-political factors, anchored the use of this quality data in fully realising its potential for improving student outcomes.

## 2　Literature Review

Schools within the United States for more than a decade have been focused on using data to improve outcomes for students (Coburn and Turner 2012). This has been a result of the 'No Child Left Behind Act' (NCLB, 2001). As a result of the NCLB Act, there have been many research papers published on the use of data in education within the United States. The reported success of utilising data in these papers is generally described as limited with many identified problems related to the production of 'quality data' and its subsequent 'use' (Goren 2012; Honig and Venkateswaran 2012; Little 2012; Luo 2008; Pierce and Chick 2011). Similarly, while a large body of education studies have focused on developing quality curriculum frameworks, studies that highlight actual methods for developing the corresponding quality data frameworks are relatively few. Using a definition of data quality (Fox et al. 1994), the remainder of this literature review discusses the challenges of recording quality data through the use of quality data frameworks. 'Data quality' is described as a multidimensional construct with the dimensions of quality including accuracy, timeliness, consistency, and completeness (Fox et al. 1994; Lee Wang and Strong 2003).

Data accuracy is defined as the "measurement or classification detail used in specifying an attribute's domain" (Fox et al. 1994, p 14). In this research case, data accuracy refers to how data describes a single or series of student learning's reflecting their progress in the context of a classification schema that defines the pedagogical learning approach and framework adopted by the school. It has been incumbent on teachers to either use data from external sources to the school or to produce data themselves through the application of local measurement instruments. Both of these scenarios have proven to be relatively ineffective in producing accurate and timely data. While data collected external to schools have well developed metadata models to measure specific outcomes, these instruments have been shown to have little relevance and validity to student outcomes (Ikemoto and Marsh 2007; Marsh et al. 2006). In contrast, data collected by school-based personnel is often limited in quality due to the lack of skills, time and organisational structures to effectively produce and use data (Bernhardt 2000). Problems with relying on teachers to collect data have been reported by Marzano (2003). Marzano states that data collected about student performances are often indirect measures with no explanatory model to interpret the data. In these cases a metadata model has not been correctly incorporated as part of the improvement program. As a result, data collected and reported by teachers is often of the wrong type or format and, therefore, is further described as irrelevant, invalid or inaccurate (Olson 2002; Rudner and Boston 2003).

From a data timeliness perspective, data in the education context should be captured as quickly as possible after the student's attempt at a learning activity so that it can be available as a feedback and analysis tool. Various authors, however, discuss that the frequency of measures available that define improvements from the input/benchmark to the output, are too low (Choppin 2002; DeLoach 2012; Hanks 2011; Marsh 2006) for example, reported that in general teachers preferred the use of classroom data to periodic external exams, stating that external exams did not provide useful data in a timely fashion. Teachers could not act on this data, as students had already moved to another teacher and or grade level. "For this reason many districts and schools rely on local tests that are issued more frequently throughout the year, thus, providing diagnostic information that could be acted on immediately" (Marsh 2006, p 114). Historically, the problem with relying on teachers to collect data is that such a process is resource intensive and, therefore, is unattractive as part of any long term data strategy. The infrequent collection about a student's progress leads to problems associated with data inconsistency.

Data consistency refers to the "probability that an item will perform a required function under stated conditions for a stated period of time" (Fox et al. 1994, p 15). In order for reported data to be considered consistent, the data collection process should be stable and consistent across collection points and over time. Progress toward student learning goals should reflect real changes rather than variations in data collection approaches or methods. Data consistency remains the biggest challenge to generation of quality data, particularly in the secondary school context. Students have multiple teachers across several subjects across and across year levels. Variations in collection frequency as well as variations in subjective evaluations of student's progress leads to inconsistent data and, therefore, reduces the validity and relevancy of the data to the quality management program.

Finally, data completeness refers to the "degree to which a data collection has all the attributes of all entities that are supposed to have values" (Fox et al. 1994, p 15). The data requirements that describe student learning progress should be clearly specified based on the information needs of the school and defined by their pedagogical framework. Data collection processes should be developed to capture the entities required in order to evaluate the progress of students with respect to the student's needs in achieving outcomes with respect to the relevant pedagogical framework.

Education researchers have cited the increasing need for improved information systems with data storage and data retrieval capacity. The ability to present the data in formats that are meaningful to school leaders and teachers has been emphasised (Rudner and Boston 2003). Although technology may be available, school districts often lack the funds or do not allocate the resources necessary to establish coherent and high-level data system capability (Olson 2002). In order to develop such technology many considerations with respect to organisational requirements and sociotechnical barriers need to be considered. Identifying the exact requirements for such an information system, as well as the exact nature of how and why barriers to use exist, is complicated. Finding a solution to these problems can be even more difficult. Within the literature review, these types of problems are referred to as wicked problems. Buchanan (1992) (citing Rittel and Webber 1973) define a wicked problem as class of social system problems which are ill-formulated, where the information is confusing, where there are many clients and decision makers with conflicting values and where the ramifications in the whole system are thoroughly confusing. From a research perspective, wicked problem types are often addressed using a Design Science Research (DSR) methodology. DSR is class of Information Systems' (IS) research that is particularly well suited for identifying, designing, developing, instantiating and evaluating solutions to wicked problems (Gleasure 2013; Hevner et al. 2004).

## 3  Research Questions

Using the DSR method, this study designs, develops, instantiates and evaluates an artefact that specifically addresses a number of education based problems associated with the capture of quality data and its use. Three research questions were developed for this study and are stated as:

**RQ1.** Did stakeholders engage with the artefact in the classroom?
**RQ2.** What was the impact of the artefact?
**RQ3.** How was data perceived and used as a tool for improving student behavior management?

Through the measurement and evaluation of these three research questions, Design Theory for the design and development of classroom based education systems is forwarded in section 8.

## 4  Methodology

The research design adopted for this study is classed as a mixed methods procedure with Design Science Research (DSR) as its main methodological approach. A Design Science methodology, however, can consist of a number of further varying research techniques. These techniques investigate both design and natural science phenomena. Davis and Olsen (1995) argue that IT research is situated within both the design and natural sciences and both research paradigms are needed for effective IT research. Further studies have supported this notion (Lee 1999; Lee et al. 1999; March & Smith 1995).

This study, therefore, uses both qualitative and quantitative research techniques to investigate the quality of artefact design, as well as resultant sociotechnical interaction with the artefact. Three distinct cycles of the DSR method are used in this study; the relevancy, design, and rigor cycles (Hevner et al. 2004). This majority of the steps from Alturki et al. (2011) DSR roadmap are used within each of the research cycles. The methods used and the outputs for each of these cycles for this research project are briefly described over the next three sections.

# 5 Relevance Cycle

The method for determining understanding this wicked problem and the development of the resultant solution are the same for this research (Rittel and Webber 1973). This research decomposes the wicked problem using Enterprise Architecture (EA) methods. Once the problem is decomposed, solutions to the parts of the decomposed problem become inputs at each architecture layer (strategy, business, application, technology layers) of the artefact. Additional inputs to the architecture design are gained through: 1. interviewing end users about the quality issues with existing classroom based IS'; 2. analysis of the existing SQL data quality captured by the legacy classroom based information systems, and; 3. requirements based on best industry practices at each architectural layer of the Enterprise. The activities of the relevance cycle are described in Table 1. Once the wicked problem had been identified and modelled, this research then initiated the Design Cycle.

| Definition Formulation Step | Description |
| --- | --- |
| 1. Determine Problem Domain | Problem is first defined as a relationship between either the Person, Component, System or Environment domains |
| 2. Set Wickedness Definition | Focus for the wicked definition was determined by types of problem domains being investigated. |
| 3a. Describe Entities | All relevant entities are described in each identified domain. |
| 3b. Describe Entity Relationships | For relevant entities, the relationship between entities within domain and between domains are described. |
| 4. Describe the misaligned or missing entities and their relationships | The wicked problem is fully described according to the misaligned or missing entities in the context of all relevant entities. |
| 5. Validate the problem entities with multiple sources | Multiple sources and validation techniques are used to triangulate the real source of the problem. |

*Table 1. The steps undertaken as part of the relevance cycle in this study*

As a result of undertaking the relevance cycle, a number of inputs and considerations are identified for future development of the artefact. Most critically the key for the success for the development of artefact was the ability to use the IS within the classroom. This requirement resulted in the development of a Bluetooth notification systems that would automate most of user interactions associated with data entry and data navigation based on the teacher's proximity to the student. Through this design the number of interactions required to capture data and find data about the individual student was significantly reduced in comparison to the legacy IS. The conceptual prototypes of this novel artefact are described in detail in the next section.

# 6 Design Cycle

As part of the relevance cycle the various entities of the problem and its relationship to the environment in which it is examined was documented. As part of the Design Cycle in this study, a review of the dependencies and co-dependencies between each of the entities was further completed. The problem entities, or group of entities that were identified as having the greatest number of dependencies, became the initial focus for our design. A cascade approach for determining the importance of the entities and the priority in which they should be addressed was undertaken as part of this approach. Through using this process, it was believed that the core of the wicked problem was accounted for, with all other dependencies appropriately documented and included as considerations in the design and development of the novel artefact. Additionally, through decomposing the wicked problem into its base entities (important to note that standard TOGAF v9.1 language was used to avoid confusion), potential solution pathways could be identified by iterating through those entities and those identified relationships that were misaligned. This iteration process was seen as a creative exercise, determining how a new artefact may work across the various entities and how it might affect their relationships.

For each iteration, the researcher identified possible uses of technology that could be utilised for each pathways and documented its applicability and limitations. A review of the solution pathways was then conducted to review whether there was a need to develop novel solutions to the identified focus problems. Given that a clear understanding of the scope and structural requirements for the artefact, is gained through this process, an exploration period began to identify potential technology adaptions to meet the specific problem identified. This led to the artefact prototyping stage.

## 6.1 Prototype Development and Testing

Two types of prototypes were developed for this study: 1. the conceptual prototype, and; 2. the minimum working product (novel component only). The conceptual prototype was initially socialised with a number of university and industry professionals to test for its conceptual viability. The conceptual prototype is shown in Figure 1.  Once a conceptual prototype had been established, the concept moved to the prototyping stage (minimum working product). Figure 2 shows a screenshot of the initial prototype that was developed, to test the conceptual prototype. This prototype was socialised with the potential end users of the IS (with a bit of imagination) to discuss possibilities.

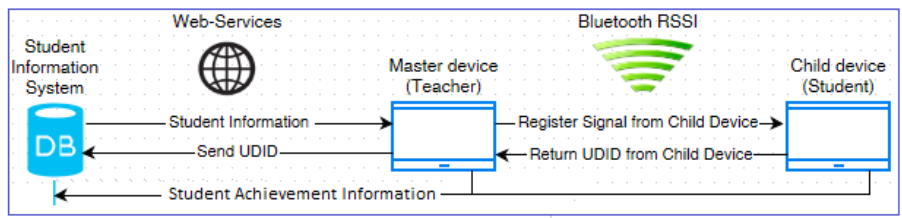

*Figure 1. The conceptual prototype of the Bluetooth signal based automated look-up function.*

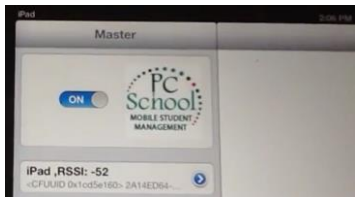

*Figure 2. The prototype developed for the study tested the viability of the conceptual principle. The figure shows the mobile device capturing multiple other mobile device ID's and their Received Signal Strength Indication (RSSI) strength.*

The final stage of the development cycle was to fully develop the artefact for use and testing. The final artefact is described in section 6.2 and 6.3.

## 6.2 Description of Novel Artefact

The novel artefact is described as a 'Teacher app' and a 'Student app' with the ability to detect multiple signals from multiple iPADs. Both the teacher and student apps read and write data to and from a Student Information System (SIS) using web services. As part of the Bluetooth Low Energy (BTLE 4.0) framework in iOS, these apps are often referred to as a Master App and a Child App. The BTLE 4.0 framework is set of Objective-C 'methods' that allows multiple Slave devices (in this case student iPads) to be detected and 'paired' to a Master device, through the Bluetooth signal.

Once the Teacher app and the Student app have 'connected' small bits of information between the two apps can be exchanged. The framework is not designed to allow for large data streaming between apps but rather the communication of small bits of information. In this case the Master device, receives two bits of information from the Slave device; the Unique Device Identifier (UDID) and its Bluetooth signal strength. When the Slave app is first used, the UDID (code generated) is written to the SIS. The UDID is then used by the Master app to automatically obtain student data from the SIS. Through this process, much of the work that is normally associated with data entry is automated. Using Bluetooth signal, the UDID of the closest student iPAD is used to automatically look up the students' details. Student information and progress information from the SIS is automatically displayed on the teacher app when the teacher is in proximity to the student. The number of interactions required by teachers using this model is reduced to one third of the interactions that was required by the legacy IS.

In addition to this novel functionality, the app includes all of the requirements documented in the Enterprise Information Architecture document i.e., the requirements to realise the service strategy, enable business functions and to ensure the applications and technology layers have the correct functional design. The various screens for the mobile based application is shown in Figure 3.

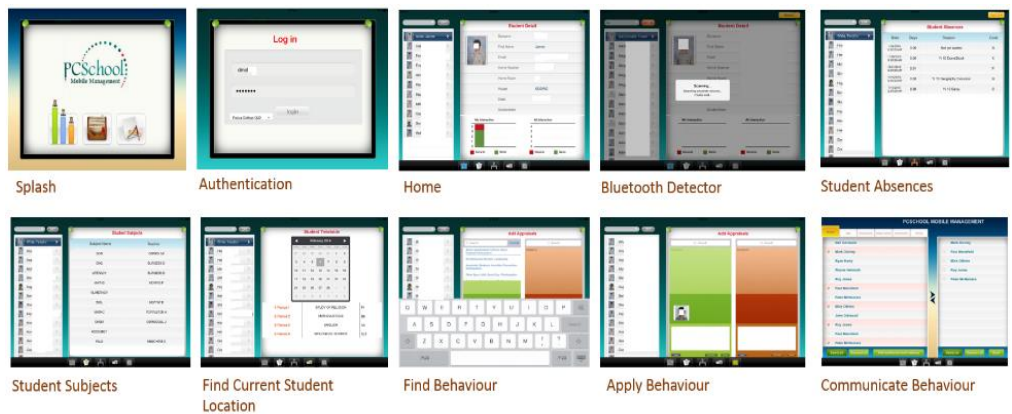

*Figure 3. The final artefact containing the novel technology component. The data is pushed and pulled through a Bluetooth notification system.*

## 6.3   Designing for Quality Data throughout the Data Cycle.

A further key requirement for the design of the software centred on the need to ensure quality data and its use throughout the continuous improvement cycle. The app is carefully engineered to ensure that each of the elements that define quality data is addressed, namely data accuracy, data consistency, data completeness and data timeliness (Fox et al. 1994; Lee Wang and Strong 2003). Quality data is then made available throughout the 'continuous improvement cycle' for the service defined. Figure 3 (Home Screen) above, for example, highlights the realisation of the continuous improvement cycle through the use of 'live' reports on the home screen. The first graph indicates the teacher's alignment with the business strategy. The second graph shows the KPI's for how the school is tracking overall with their approach to achieving the service strategy.

Once the final artefact was complete it was ready to trial within an environment and tested for its efficacy. In the Design Science Roadmap, both 'Artificial' and 'Naturalistic' testing methods were conducted within the Design Cycle. For the purposes of this study the artificial testing was conducted as part of the Design Cycle and the Naturalistic testing was conducted as part of the Rigor Cycle.

# 7   Rigor Cycle

To evaluate and communicate the sociotechnical success of the artefact, the rigor cycle for this study consisted of three parts: 1. the sociotechnical interaction was first measured using the instruments specified in the Data Analysis Techniques section; 2. a discussion of these results examining the effectiveness of the artefact in meeting the objectives of the research is then completed, and; 3. Design Theory for this research is finally made explicit.

## 7.1   Data Analysis Techniques

Three instruments are used to measure the sociotechnical effect of the artefact. The Unified Theory of Acceptance and Use of Technology (UTAUT) scale was used to measure the user's engagement with the artefact (Venkatesh et al. 2003). The IS-Impact Scale was used to measure System Quality, Information Quality, Individual Impact and Organisational Impact (Gable et al. 2008). Finally a convergent interview technique was used to gain a qualitative understanding of the interaction between the user and environment and the resultant data use (Jepsen and Rodwell 2008). For each research question in this study, the scale, analysis technique and results are shown in Appendix 1. In this study, Cronbach's coefficient alphas, which are calculated based on the average inter-item correlations, were used to measure internal consistency. Overall, the result shows that all alpha values of the instruments used, were reliable and exhibited appropriate construct reliability.

## 7.2   Data Collection Sites

All surveys were distributed and applied within a single site. The quantitative surveys were applied pre and post instantiation of the artefact. The convergent interview process was conducted after the evaluation trial period of the artefact had concluded.

## 7.3　Population, Sample and Research Period

Surveys were distributed to both teaching staff and teaching support staff at the college. The total population of teachers at the college was ninety-four (94). In order for the subject's responses to be valid, a single user was required to successfully complete both pre and post questionnaires. Table 4 (in the Appendix) summarises the survey, showing the research hypothesis tested, the scale origin, the analytical methods used in evaluating responses, and the findings. The sample size for both the UTAUT and IS-Impact questionnaires was 32. This represented 38.09% of the total teaching staff at the college. A total of 12 respondents completed the convergent interviews post implementation of the artefact. This represented 14.28% of the total teaching staff at the college. Data was collected during the first term of 2014 and the third term of 2014. These two periods corresponds to the pre and post periods of artefact implementation. Given the frantic nature of schools at the beginning and end of terms, it was discerned that all measurements would be best applied between week three and seven of term. This time period was chosen to ensure that there was no overlap with the marking / reporting period for teachers.

## 7.4　RQ 1 (Teacher engagement with the artefact)

The results for research question 1 (RQ1) showed that significant differences between UTAUT measures pre and post application of the artefact existed for the constructs of; performance expectancy, effort expectancy, facilitating conditions, and use. Social influence and behavioural intention were not different pre and post application of the artefact. The results are shown in Table 2.

| UTAUT Construct | Mean 1 | Mean 2 | Z statistics | Sig (2-tailed) |
|---|---|---|---|---|
| Performance expectancy | 9.67 | 15.62 | −4.085 | 0.001* |
| Effort expectancy | 11.75 | 16.63 | −3.954 | 0.001* |
| Social influence | 12.00 | 11.15 | −0.604 | 0.557 |
| Facilitating conditions | 14.14 | 14.62 | −2.380 | 0.016* |
| Behavioural intention | 17.57 | 12.00 | −1.338 | 0.186 |
| Use | 9.88 | 16.35 | −28.34 | 0.003* |

*Table 2. Statistical outcomes of the UTAUT scale　　* Statistical Significance*

## 7.5　RQ 2 (Impact of the artefact)

The impact of the artefact was measured through the application of the IS-Impact scale as well as the direct examination of the data quality in the SQL database attached to the artefact. The results for the IS-Impact scale are shown in Table 3.

| Construct | Mean | SD | SEM | Lower | Upper | t | df | Sig. |
|---|---|---|---|---|---|---|---|---|
| Individual impact | −2.37 | 3.42 | 0.60 | −3.60 | −1.14 | −3.92 | 31 | 0.001 |
| Organisational impact | −2.68 | 5.97 | 1.05 | −4.83 | −0.53 | −2.54 | 31 | 0.016 |
| Information quality | −4.65 | 5.97 | 1.05 | −6.80 | −2.50 | −4.41 | 31 | 0.001 |
| System quality | −9.71 | 7.23 | 1.27 | −12.32 | −7.10 | −7.59 | 31 | 0.001 |
| Satisfaction | −7.71 | 8.14 | 1.44 | −10.65 | −4.78 | −5.35 | 31 | 0.001 |

*Table 3. Statistical outcomes of the IS-Impact scale　　* Statistical Significance*

The results of the IS-Impact scale showed that the instantiation of the artefact had a significant and positive impact on the constructs: individual impact, organisational impact, information quality, system quality and satisfaction.

The results of the SQL analysis highlighted the SQL data use patterns associated with the new artefact. These 'use patterns' were compared to the same period of the previous year, thus providing an insight into the effect of the trial artefact. Figure 4, below shows a sample of data prior and post instantiation of the artefact. Data quality improved with respect to its accuracy, consistency and completeness.

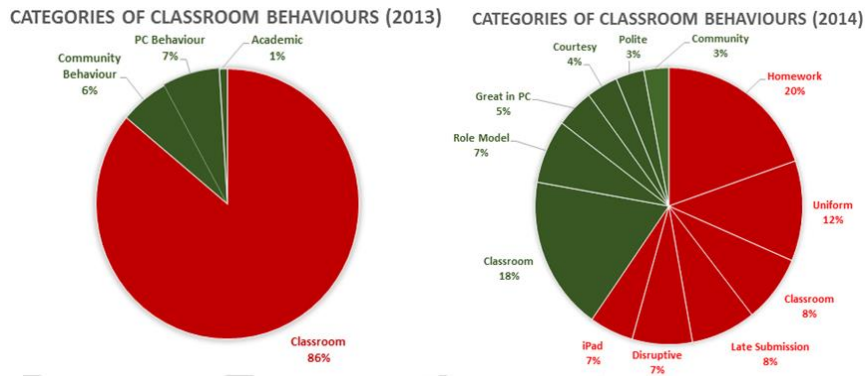

*Figure 4. Example of data quality pre and post instantiation of the artefact.*

### 7.6　RQ3 (How was data perceived and used as a tool for teaching)

Interview data was collected to further understand why, or why not, the artefact was accepted by end-users and this data provided a rich context to the user's engagement levels with the artefact. A number of questions were asked in the convergent interviews eliciting content such as: 1. the role of data and IS' in education; 2. how data is currently used to inform practice; 3. engagement issues with the use of information technology as a whole; 4. engagement issues with information systems and data; 5. the quality of data they are currently exposed to; 6. the quality of the legacy IS; 7. the quality of the newly instantiated artefact; 8. the correlation between IS quality and their reporting behavior, and; 9. the relationship between teacher behavior and student outcomes.

The results of the convergent interviews revealed, that while teachers generally gave positive feedback with respect to the instantiated artefact, exogenous factors to the artefact such as: 1. The school does not have a data-driven culture; 2. mistrust of why and how data is used; 3. complexity of reporting requirements, and; 4. Habit, were a major factors in anchoring the use of the artefact on a larger scale.

## 8　Discussion

The aim of this study was to further knowledge of information systems design within the education context. Contrary to previous studies on data use within education, this study, through the use of a DSR methodology forwards insight into the complex internal mechanisms of data use within education. This study has provided effective methodologies for improving the quality of IS' within the education context. In this section, discussion on how design improvements of the IS, leads to quality of information throughout the continuous improvement cycle, is forwarded. This section provides insight into those endogenous and exogenous variables that influence the various stages of the continuous improvement cycle for the behaviour management service.

The artefact design approach taken was complex. First the purpose and function of the IS was addressed. The design methodology was detailed in the design cycle section, and the artefact is defined as the sum of the requirements defined in the Enterprise Information Architecture (EIA). Once the artefact had been designed from an architectural perspective, further design considerations were made to ensure that quality information was available throughout the defined continuous improvement (CI) cycle. Figure 5 shows the data/information cycle developed and included as part of the artefact' design.

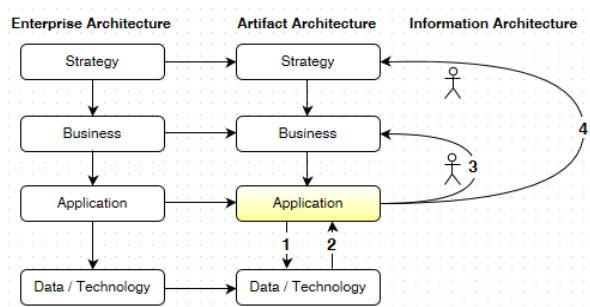

*Figure 5. Information flow paths as part of the continuous improvement cycle.*

Figure 5 shows the information flow paths that forms the CI cycle for this service. The CI is described as: (1) the user writes data via the application to the data stores; (2) the user also receives feedback about the quality of this data; (3) the user (both teacher and student) receives feedback at an application level about their work behaviour in terms of aligning it to best practice standards, and (4) the user receives feedback from the application about whether their actions align with the organisational strategy for this particular business service.

The unique environmental factors hindering the usability of IS, and subsequent data use in the classroom, were documented as part of this study. Teachers strongly argued it was impractical to enter data in the classroom while teaching. This research documented several issues in regards to this problem (highlighted in Figure 5.0 as flow path 1). A novel technology was developed to improve the ability of teachers to use the artefact within the classroom environment. Through this, it was believed that teachers could be empowered to both capture/record and use the quality data within the classroom. This specific design of the artefact addressed the information flow paths (1 & 2).

In relation to flow paths (1 & 2), the results of the UTAUT questionnaire found that through the instantiation of the artefact that the mean scores for effort expectancy decreased, however, scores for performance expectancy, facilitating conditions and use significantly increased. Behavioural intent and social influence were not significantly different pre and post instantiation of the artefact. This study has concerns in regards to the predictive and nomological validity of the behavioural intention construct used within this study.

The results from the IS-impact scale indicated that the artefact did have a positive impact on end-users. There were, however, no between-group effects (role type) for 'individual impact'. It was expected that the impact of the artefact would be different between teachers and administrators who use data. The results also showed that information quality and system quality were perceived to have improved with the implementation of the new artefact. The SQL data analysis supported the finding that information quality had improved. The SQL data clearly showed an improvement in the range and type of data entered into the database as a result of the instantiation of the new artefact. The comparison of entered row counts of data between the legacy IS and the instantiated artefact were double in nearly all cases. This study can draw definite conclusions that the artefact design method led to improved artefact quality and subsequently improved data quality.

The design of the new artefact also specifically included functionality that enabled teachers to exercise best-practice behaviour management in the classroom. This functionality provided immediate feedback about teacher's own actions according to this 'best practice'. This design addressed information flow path (3). Finally, the apps also provided comparison data (information) on teacher's behaviour management practice in relation to that of other teachers at the school. A major part of behaviour management requires that students receive consistent feedback on a behaviour they are exhibiting from their seven teachers. Inconsistent feedback frustrates teachers and students and diminishes and reinforcement strength. This functionality is addressed by information flow path (4).

By satisfying the four information flow paths identified in Figure 5, it was expected that data and information would be perceived as accurate, relevant and timely and, therefore, would be used to inform and improve practice. The results from the convergent interviews, however, showed that exogenous factors acted as barriers to teachers using this improved quality of data to inform teacher practice. This study cannot definitively determine reasons for the interaction between users and information flow paths (3) and (4). The results from the convergent interviews suggested that teachers were more likely to continue their normal habits, even with direct feedback suggesting they were not aligning their work habits with best behaviour management practices.

As per the attitude-behaviour management model (Azjen 1980; Azjen and Fishbein 1991), this study found that the use of data throughout the CI cycle is anchored by exogenous variables, such as organisational habit and culture. It was reported in the convergent interviews that there was not a culture of using data to inform teacher practice. It was shown that teachers typically did not engage with the use of data for many of the reasons, and these were similar to those highlighted in the literature review. Many teachers at the application domain did not have an explicit understanding of behaviour management principles. They, therefore, were unable to perceive the value of the artefact with the incorporated behaviour management functions. It was shown in the convergent interviews that teachers did not make the link between artefact quality and student behavioural outcomes and, therefore, judged the artefact quality according to its utility to make their role easier. This was evidenced according to effort expectancy, and performance expectancy *mean* scores on the UTAUT scale. To overcome the issues of habit and culture as barriers to IS use, teachers require a greater understanding of behaviour management principles and will need to be made explicitly aware of the direct effect their appraisal behaviour has on the student (consequences of behaviour).

## 9 Conclusion

The purpose of this study was to design, develop, instantiate and evaluate an IS artefact in order to facilitate the continuous collection and its use in the education sector. A DSR methodology was used in this research. The relevance cycle within the DSR method, used specific methods to define, and classify the research problem. The design cycle adopted the design methodology suggested by Alturki et al. (2011) to build the artefact. The UTAUT, IS- Impact and Convergent Interview techniques were used to evaluate the socio-technical response to the instantiation of the artefact. The results of the evaluation showed that the artefact had a significant effect on perceived quality of the artefact. It was also perceived to have improved information quality. The improved quality of data, however, was not effectively used to improve the alignment of work practices to 'best business practice' and the service strategy. Many exogenous factors limited the use of this improved data quality.

### 9.1 Limitations to the study

This study was situated in one co-educational school that spanned years 5 to 12. This school had a well-established pastoral care / behavior management model and a focus on increasing student engagement with the full spectrum of its curricular and co-curricular programs. Though all schools within the Australian educational environment have a focus on improved student behaviour, not all have the resources and professional development programs that support its effective implementation. Most schools do not have an integrated suite of information systems that accurately capture both positive and negative student behaviours. The staff uptake of this reported IS suite may be a function of this particular school environment, and a broader set of schools using this application suite is required to more fully test user acceptance and the effects on student behaviour and teacher reporting patterns of student behaviour. A major limitation to this research is the construct validity of UTAUT in the absence of volitional information systems use. Modifications to the scale are required in research where information systems are trialed for a set period of time.

### 9.2 Recommendations for future research

This study showed that information systems quality impacted student behaviour management, user acceptance and data quality metrics. Further work is required to determine if the increase in positive behaviour leads to changes in the perception of student behaviour by the teachers, as well as increased engagement in the learning process by students. The effect of the efficiencies of the new behaviour management system on teacher teaching time should be examined. The impact of 'intention to use' in UTAUT needs further analysis, given that teachers are required to monitor and input behaviour data in the classroom and co-curricular program. This longitudinal set of studies will be the focus of the research program in 2016.

## Appendix 1

|  | Research Question / Hypothesis | Scale | Analysis | Supported |
|---|---|---|---|---|
| **RQ 1** | **Does the specific IS design lead to improved engagement with the** |  |  | **There was evidence to support that IS design did lead to improved engagement** |
| h1 | The new artefact will positively influence teacher's intention to use it. | UTAUT | t-tests, ANOVA | Could not be supported. Volitional issues with the UTAUT questionnaire and its |
| h2 | The measures of UTAUT mediate teacher's intention to use the new artefact. | UTAUT | Pearson's Correlation, t-tests | PE, EE, SI, FC had relationships to USE. PE, EE, SI, FC different pre and |
| H3 | The new artefact design will have an impact on the individual. | UTAUT | ANOVA | Individual Impact significantly different pre and post- test. |
| H4 | The new artefact design will lead to increased use. | UTAUT | Descriptive Statistics, t-tests | Significant increase in row counts (SQL Analysis) |
| **RQ2** | **What was the impact of the newly instantiated artefact?** |  |  | ***The evidence shows Significant and positive impact on SQ, IQ, II and OI*** |
| h5 | The new artefact will improve perceptions about the System and Information Quality. | IS-Impact | Pearson's Correlation, | Both System Quality (SQ) and Information Quality (IQ) were significantly different pre and post-test |
| h6 | The new artefact will have a positive Impact on the Individual and the Organisation. | IS-Impact | Pearson's Correlation, t-tests | Constructs for Individual Impact (II) and Organisational Impact (OI) were significantly different pre and post-test |
| h7 | The new artefact will improve the quality of data measuring student learning behaviours. | IS-Impact | Descriptive statistics | Data suggests that there was improvement across all quality attributes |
| **RQ3** | **How was data perceived and used as a tool for improving teacher practice?** |  |  | **Generally the tool itself was seen as positive, however, its value was diminished in the light of exogenous** |
| h8 | Teachers will perceive the artefact has having utility for their role. | Convergent Interviews | Interview Data technique | The artefact was generally viewed as having positive utility for the classroom |
| h9 | Teachers will use the artefact uninhibited by exogenous factors to the artefact. | Convergent Interviews | Interview Data technique | Results suggest that exogenous factor to the IS inhibited its use. |
| h10 | Stakeholders will perceive a positive relationship between artefact quality and their reporting behaviours. | Convergent Interviews | Interview Data technique | There was mixed feedback in the interviews with regards to this topic. |
| h11 | Teachers will perceive a positive relationship between their reporting behaviours and student outcomes. | Convergent Interviews | Interview Data technique | There was mixed feedback in the interviews with regards to this topic. |

*Table 4. Results and outcomes for applied evaluative data techniques*
Performance Expectancy (PE), Effort Expectancy (EE), Social Influence (SI), Facilitating Conditions (FC) Individual
Impact (II), Organisational Impact (OI), System Quality (SQ), Information Quality (IQ)